\documentclass[nofootinbib,floats,aps,superscriptaddress,preprint,tightenlines]
{revtex4}

\def\D0bar{\overline D{}^0}
\def\K0bar{\overline K{}^0}
\def\etackm{\eta_{\rm CKM}}
\def\etacp{\eta_{CP}}
\def\3bar{\overline{3}}
\def\sixbar{\overline{6}}
\def\tenbar{\overline{10}}
\def\15bar{\overline{15}}
\def\24bar{\overline{24}}
\def\42bar{\overline{42}}
\def\60bar{\overline{60}}
\def\cO{{\cal O}}
\newcommand{\BR}{{\cal B}}
\newcommand{\beq}{\begin{equation}}
\newcommand{\eeq}{\end{equation}}
\newcommand{\beqa}{\begin{eqnarray}}
\newcommand{\eeqa}{\end{eqnarray}}

\begin{document}

\preprint{\vbox{\hbox{JHU--TIPAC--2001-04} \hbox{LBNL--48969}
     \hbox{WSU--HEP--0101} \hbox{hep-ph/0110317}\hbox{October, 2001}}}

\vspace*{.75in}

\title{$SU(3)$ Breaking and $D^0-\D0bar$ Mixing}

\author{Adam F.\ Falk\vspace{10pt}}
\affiliation{Department of Physics and Astronomy,
      The Johns Hopkins University\\
      3400 North Charles Street, Baltimore, MD 21218\vspace{6pt}}

\author{Yuval Grossman}
\affiliation{Department of Physics, Technion--Israel Institute of 
Technology\\
      Technion City, 32000 Haifa, Israel\vspace{6pt}}

\author{Zoltan Ligeti}
\affiliation{Ernest Orlando Lawrence Berkeley National Laboratory\\
       University of California, Berkeley, CA 94720\vspace{6pt}}

\author{Alexey A.\ Petrov}
\affiliation{Department of Physics and Astronomy\\
     Wayne State University, Detroit, MI 48201\\ $\phantom{}$ }

\begin{abstract}
The main challenge in the Standard Model calculation of the mass and
width
difference in the $D^0-\D0bar$ system is to estimate the size of $SU(3)$
breaking
effects.  We prove that $D$ meson mixing occurs in the Standard Model 
only at second order
in $SU(3)$ violation. We consider the possibility that phase space 
effects may be the
dominant source of $SU(3)$ breaking.  We find that 
$y=\Delta\Gamma/2\Gamma$ of the order of one
percent is natural in the Standard Model, potentially reducing the 
sensitivity to new physics of measurements of $D$ meson mixing.
\end{abstract}

\maketitle

\section{Introduction}

It is a common assertion that the Standard Model prediction for mixing
in the
$D^0-\D0bar$ system is very small, making this process a sensitive probe
of new
physics.  Two physical parameters that characterize $D^0-\D0bar$ mixing
are
\beq
x \equiv {\Delta M \over \Gamma}\,, \qquad
    y \equiv {\Delta \Gamma \over 2\Gamma}\,,
\eeq
where $\Delta M$ and $\Delta \Gamma$ are the mass and width differences
of the
two neutral $D$ meson mass eigenstates, and $\Gamma$ is their average
width.
The $D^0-\D0bar$ system is unique among the neutral mesons in that it is
the
only one whose mixing proceeds via intermediate states with down-type
quarks.
The mixing is very slow in the Standard Model, because the third
generation
plays a negligible role due to the smallness of $|V_{ub} V_{cb}|$ and 
the relative smallness of $m_b$, and
so the
GIM cancellation is very effective~\cite{Ge92,Oh93,ap,Da85,dght}.

The current experimental upper bounds on $x$ and $y$ are on the order of 
a few times
$10^{-2}$, and are expected to improve significantly in the coming
years.  To
regard a future discovery of nonzero $x$ or $y$ as a signal for new 
physics, we
would
need high confidence that the Standard Model predictions lie
significantly
below the present limits.  As we will show, in the Standard Model $x$ 
and $y$ are generated
only at second
order in $SU(3)$ breaking, so schematically
\beq
x\,,\, y \sim \sin^2\theta_C \times [SU(3) \mbox{ breaking}]^2\,,
\eeq
where $\theta_C$ is the Cabibbo angle.  Therefore, predicting the
Standard Model
values of $x$ and $y$ depends crucially on estimating the size of
$SU(3)$ breaking.  Although $y$ is expected to be determined
by
Standard Model processes, its value nevertheless affects significantly 
the sensitivity to new physics of experimental analyses of $D$ 
mixing~\cite{BGLNP}.

At present, there are three types of experiments which measure $x$ and
$y$.  Each is actually sensitive to a combination of $x$ and $y$,
rather than to either quantity directly.  First, there is the $D^0$
lifetime difference to $CP$ even and $CP$ odd final 
states~\cite{e791y,focusy,cleoy,belley,babary}, which to leading order
measures
\beq\label{ycp}
y_{CP} = {\tau(D \to \pi^+ K^-) \over \tau(D \to K^+ K^-)} - 1
    = y\cos\phi - x \sin\phi\, \frac{A_m}2\,,
\eeq
where $A_m = |q/p|^2-1$ (see Eq.~(\ref{definition1}) for the definition 
of the neutral $D$ mass eigenstates), and $\phi$ is a possible $CP$ 
violating phase of the
mixing amplitude.
Second, one can measure the time dependence of doubly Cabibbo
suppressed decays,
such as $D^0\to K^+ \pi^-$~\cite{Godang:2000yd}, which is sensitive to 
the three quantities
\beq\label{xyprime}
(x \cos\delta + y \sin\delta) \cos\phi\,, \qquad
    (y \cos\delta - x \sin\delta) \sin\phi\,, \qquad
    x^2 + y^2\,,
\eeq
where $\delta$ is the strong phase between the Cabibbo allowed and doubly
Cabibbo suppressed amplitudes.  A similar study for  $D^0\to 
K^-\pi^+\pi^0$ also would be valuable, with the strong phase difference 
extracted simultaneously from the Dalitz plot analysis~\cite{brand}.  
Third, one can search for $D$ mixing in semileptonic 
decays~\cite{DmixSL}, which is
sensitive to $x^2+y^2$.

In a large class of models, the best
hope to
discover new physics in $D$ mixing is to observe the $CP$ violating
phase,
$\phi_{12} = \mbox{arg}\, [M_{12}/\Gamma_{12}]$ (see the definitions 
(\ref{del-mass}) and (\ref{del-width}) below), which is very small in
the
Standard Model.  However, if $y \gg x$, then the sensitivity of any
physical
observable to $\phi_{12}$ is suppressed, since $A_m$ is proportional to 
$x/y$ and
$\phi$ is to $(x/y)^2$, even if new physics makes a large contribution 
to $\Delta M$~\cite{BGLNP}.   It is also clear from Eq.~(\ref{xyprime})
that if $y$ is significantly larger than
$x$, then $\delta$ must be known very precisely for experiments to be
sensitive to new physics in the terms linear in $x$ and $y$.  It may be
possible to measure $\delta$ with some accuracy at the planned 
$\tau$-charm factory CLEO-c~\cite{Silva:2000bd,Gronau:2001nr}.

There is a vast literature on estimating $x$ and $y$ within and beyond
the
Standard Model; for a compilation of results, see Ref.~\cite{nelson}.
Roughly, there are two approaches, neither of which give very reliable
results because $m_c$ is in some sense intermediate between heavy and
light.  The ``inclusive'' approach is based on the operator
product expansion (OPE).  In the $m_c \gg \Lambda$ limit, where 
$\Lambda$ is a scale characteristic of the strong interactions, $\Delta 
M$ and
$\Delta\Gamma$ can be expanded in terms of matrix elements of local
operators~\cite{Ge92,Oh93,Bigi:2000wn}.  Such calculations yield $x,y
\lesssim
10^{-3}$.  The use of the OPE relies on local quark-hadron duality, and
on
$\Lambda/m_c$ being small enough to allow a truncation of the series 
after
the
first few terms.  The charm mass may not be large enough for these to be
good approximations, especially for nonleptonic $D$ decays.
An observation of $y$ of order $10^{-2}$ could be ascribed to a
breakdown of the OPE or of duality~\cite{Bigi:2000wn},  but such a large 
value of $y$ is certainly not a generic prediction of OPE analyses.  
The ``exclusive'' approach sums over intermediate hadronic
states,
which may be modeled or fit to experimental
data~\cite{dght,Buccella:1996uy,Golowich:1998pz}.
Since there are cancellations between states within a given $SU(3)$
multiplet,
one needs to know the contribution of each state with high precision.
However, the $D$ is not light enough that its decays are dominated
by a few final states.  In the
absence of sufficiently precise data on many decay rates and on strong
phases,
one is forced to use some assumptions.  While most studies find $x,y
\lesssim
10^{-3}$, Refs.~\cite{wolf,cnp,kaeding} obtain $x$ and $y$ at the
$10^{-2}$ level by arguing that $SU(3)$ violation is actually of order
unity, but
  the source of the large $SU(3)$ breaking is not made
explicit.

In this paper, we compute the contribution to $\Delta \Gamma$ from
$SU(3)$
breaking from final state phase space differences.  This is a calculable
source of $SU(3)$
violation, which enhances the rates to final states containing fewer
strange quarks.  In Sec.~\ref{sec:form} we review the
formalism of $D^0-\D0bar$ mixing.  In Sec.~\ref{sec:su3} we give a
general group theory proof that
$\Delta M$ and $\Delta\Gamma$ are only generated at second order in
$SU(3)$
breaking if $SU(3)$ violation enters these quantities perturbatively.  In Sec.~\ref{sec:estim} we discuss the estimates of $SU(3)$
breaking
using the ``inclusive" and ``exclusive" analyses, and remind the reader
of the shortcomings of each.  Our
main results are found in Sec.~\ref{sec:ps}, namely the calculation of 
$SU(3)$
breaking in $\Delta\Gamma$ from phase space effects
in two-, three- and four-body final states.
   We find that such
effects are
very important, and can
naturally account for  $\Delta\Gamma/2\Gamma$ at the percent level.  We 
extend the analysis to intermediate resonances in Sec.~\ref{sec:res}.  In
Sec.~\ref{sec:concl} we present our conclusions and ask whether in light 
of our results it
remains possible
for the measurement of $D$ mixing to probe new physics.

\section{Formalism}\label{sec:form}

We begin by reviewing the formalism for $D^0-\D0bar$ mixing.  The mass
eigenstates $D_L$ and $D_S$ are superpositions of the flavor eigenstates
$D^0$
and $\D0bar$,
\begin{equation} \label{definition1}
   | D_{L,S} \rangle = p\, | D^0 \rangle \pm q\, | \D0bar \rangle\,,
\end{equation}
where $|p|^2 + |q|^2=1$.    In the Standard Model $CP$ violation in $D$ 
mixing is negligible, as is $CP$ violation in $D$ decays both in the 
Standard Model and in most scenarios of new physics.  From here on we 
will assume that $CP$ is a good symmetry.  Then $p=q$, and $| D_{L,S} 
\rangle$ become $CP$
eigenstates,
\begin{equation}
    CP | D_{\pm} \rangle = \pm | D_{\pm} \rangle \,,
\end{equation}
with the mass and width differences defined as $\Delta M \equiv
m_{D_+} -
m_{D_-}$ and $\Delta \Gamma \equiv \Gamma_{D_+} - \Gamma_{D_-}$.  The
off-diagonal element of the $D^0 - \D0bar$ mass matrix can be expressed
as
\beqa
M_{12} &=& \langle \D0bar | {\cal H}^{\Delta C=2}_w| D^0 \rangle
    + {\rm P} \sum_{n}
    \frac{\langle \D0bar | {\cal H}^{\Delta C=1}_w| n \rangle
    \langle n | {\cal H}_w^{\Delta C=1} | D^0 \rangle}
    {m_D^2 - E_n^2 }\,, \label{del-mass} \\[4pt]
   \Gamma_{12} &=& \sum_{n}\rho_n
    \langle \D0bar | {\cal H}^{\Delta C=1}_w | n \rangle
    \langle n | {\cal H}_w^{\Delta C=1 } | D^0 \rangle \,,
    \label{del-width}
\eeqa
where the sum is over all intermediate states, P denotes the principal
value,
and $\rho_n$ is the density of the state $n$.  The first term in
Eq.~(\ref{del-mass}) comes from the local $|\Delta C| = 2$ operators (box
and dipenguin), which affect $M_{12}$ only.  The second term comes
from  the
insertion of two $|\Delta C|=1$ operators.  There is a contribution of 
this type to both
$M_{12}$ and
$\Gamma_{12}$.

One can then express $y$ in two equivalent ways, either as a sum over the
states that are common to $D^0$ and $\D0bar$,
\beq\label{ysum}
y = \frac{1}{2\Gamma} \sum_n\, \rho_n \left[
    \langle D^0 | {\cal H}_w | n \rangle
    \langle n | {\cal H}_w| \D0bar \rangle +
    \langle \D0bar | {\cal H}_w | n \rangle
    \langle n | {\cal H}_w | D^0 \rangle \right] ,
\eeq
or as the difference in the decay rates of the two mass eigenstates
\beq
y = \frac{1}{2\Gamma} \sum_n\, \rho_n \left[
    | \langle D_+ | {\cal H}_w | n \rangle |^2 -
    | \langle D_- | {\cal H}_w | n \rangle |^2 \right].
\eeq
A similar pair of expressions can be written for $x$,
\begin{eqnarray}
x &=& {1\over\Gamma} \left[
    \langle \D0bar | {\cal H}_w | D^0 \rangle
    + {\rm P} \sum_n {\langle D^0 | {\cal H}_w | n \rangle
    \langle n | {\cal H}_w | \D0bar \rangle
    + \langle \D0bar | {\cal H}_w | n \rangle
    \langle n | {\cal H}_w| D^0 \rangle \over m_D^2 - E_n^2}
\right],
    \nonumber  \\[4pt]
&=& {1\over\Gamma} \left[
    \langle \D0bar | {\cal H}_w | D^0 \rangle
    + {\rm P}\sum_n { |\langle D_+ | {\cal H}_w | n \rangle |^2 -
    | \langle D_- | {\cal H}_w| n \rangle |^2 \over m_D^2 - E_n^2 }
\right] .
\end{eqnarray}
Note that $x$ and $y$ are generated by off-shell and
on-shell
intermediate states, respectively.

\section{$SU(3)$ analysis of $D^0-\D0bar$ mixing}\label{sec:su3}

We now prove that $D^0-\D0bar$ mixing arises only at second order in
$SU(3)$
breaking effects.  The proof is valid when
$SU(3)$ violation enters perturbatively.  This would not be the case, for
example, if $D$ transitions were dominated by intermediate states or single resonances
close to threshold.  As we will see explicitly in Secs.~\ref{sec:ps} and~\ref{sec:res}, in such cases it is sometimes possible for $SU(3)$ violation to be enhanced substantially.
Yet other than in these exceptional situations,
treating $SU(3)$ violation perturbatively seems to us to be a mild
assumption.

The quantities $M_{12}$ and $\Gamma_{12}$ which determine $x$ and $y$
depend on
matrix elements with the general structure
\beq
      \langle\D0bar|\, {\cal H}_w {\cal H}_w\, |D^0\rangle\,,
\eeq
where in this section we let ${\cal H}_w$ denote specifically the 
$\Delta C=-1$ part of the weak Hamiltonian.  Let $D$ be the field 
operator that creates a $D^0$ meson and
annihilates a
$\D0bar$.  Then the matrix element may be written as
\beq\label{melm}
      \langle 0|\, D\, {\cal H}_w {\cal H}_w \,D\, |0 \rangle\,.
\eeq
Let us focus on the $SU(3)$ flavor group theory properties of this
expression.

Since the operator $D$ is of the form $\bar cu$, it transforms in the
fundamental representation of $SU(3)$, which we will represent with a
lower
index, $D_i$.  We use a convention in which the correspondence between 
matrix indices and quark flavors is $(1,2,3)=(u,d,s)$.  The only nonzero 
element of $D_i$ is $D_1=1$.  The
$\Delta C=-1$
part of the weak Hamiltonian has the flavor structure $(\bar q_ic)(\bar
q_jq_k)$, so its matrix representation is written with a fundamental
index and
two antifundamentals, $H^{ij}_k$.  This operator is a sum of irreducible
representations contained in the product $3 \times \3bar \times \3bar =
\15bar
+ 6 + \3bar + \3bar$.  In the limit in which the third generation is
neglected,
$H^{ij}_k$ is traceless, so only the $\15bar$ (symmetric on $i$ and $j$)
and 6
(antisymmetric on $i$ and $j$) representations appear.  That is, the
$\Delta
C=-1$ part of ${\cal H}_w$ may be decomposed as ${1\over2} 
(\cO_{\15bar} +
\cO_6)$,
where
\beqa
\cO_{\15bar} &=& (\bar sc)(\bar ud) + (\bar uc)(\bar sd)
    + s_1(\bar dc)(\bar ud) + s_1(\bar uc)(\bar dd)\nonumber\\
&&{} - s_1(\bar sc)(\bar us) - s_1(\bar uc)(\bar ss)
    - s_1^2(\bar dc)(\bar us) - s_1^2(\bar uc)(\bar ds) \,, \nonumber\\
\cO_6 &=& (\bar sc)(\bar ud) - (\bar uc)(\bar sd)
    + s_1(\bar dc)(\bar ud) - s_1(\bar uc)(\bar dd)\nonumber\\
&&{} - s_1(\bar sc)(\bar us) + s_1(\bar uc)(\bar ss)
    - s_1^2(\bar dc)(\bar us) + s_1^2(\bar uc)(\bar ds) \,,
\eeqa
and $s_1=\sin\theta_C\approx0.22$.  The matrix representations
$H(\15bar)^{ij}_k$ and $H(6)^{ij}_k$ have nonzero elements
\begin{equation}
\begin{tabular}{rll}
$H(\15bar)^{ij}_k:\qquad$
    &  $H^{13}_2 = H^{31}_2=1$\,,  &  $H^{12}_2 = H^{21}_2 = s_1$\,,\\
&  $H^{13}_3 = H^{31}_3 = -s_1$\,,  &  $H^{12}_3 =
H^{21}_3=-s_1^2$\,,\\[4pt]
$H(6)^{ij}_k:\qquad$
    &  $H^{13}_2 = -H^{31}_2=1$\,,  &  $H^{12}_2 = -H^{21}_2 = s_1$\,,\\
&  $H^{13}_3 = -H^{31}_3 = -s_1$\,,$\qquad$
    &  $H^{12}_3 = -H^{21}_3 = -s_1^2$\,.
\end{tabular}
\end{equation}
We introduce $SU(3)$ breaking through the quark mass operator ${\cal
M}$, whose
matrix representation is $M^i_j={\rm diag}(m_u,m_d,m_s)$. Although
${\cal M}$
is a linear combination of the adjoint and singlet representations, only
the 8
induces $SU(3)$ violating effects.  It is convenient to set $m_u=m_d=0$
and let
$m_s\ne0$ be the only $SU(3)$ violating parameter.  All nonzero matrix
elements
built out of $D_i$, $H^{ij}_k$ and $M^i_j$ must be $SU(3)$ singlets.

We now prove that $D^0-\D0bar$ mixing arises only at second order in
$SU(3)$
violation, by which we mean second order in $m_s$.  First, we note that
the
pair of $D$ operators is symmetric, and so the product $D_iD_j$
transforms as a
6 under $SU(3)$.  Second, the pair of ${\cal H}_w$'s is also symmetric, 
and the
product $H^{ij}_kH^{lm}_n$ is in one of the representations which
appears in
the product
\beqa
\left[ (\15bar+6)\times(\15bar+6) \right]_S &=&
    (\15bar\times\15bar)_S +(\15bar\times 6)+(6\times 6)_S \\*
&=& (\60bar+\24bar+15+15'+\sixbar) + (42+24+15+\sixbar+3)
    + (15'+\sixbar)\,. \nonumber
\eeqa
A straightforward computation shows that only three of these
representations
actually appear in the decomposition of ${\cal H}_w{\cal H}_w$.  They 
are the
$\60bar$, the
42, and the $15'$ (actually twice, but with the same nonzero elements
both
times).  So we have product operators of the form
\beqa
    DD &=& {\cal D}_6\,, \nonumber \\
    {\cal H}_w {\cal H}_w &=& \cO_{\60bar}+\cO_{42}+\cO_{15'}\,,
\eeqa
where the subscript denotes the representation of $SU(3)$.

Since there is no $\sixbar$ in the decomposition of ${\cal H}_w{\cal 
H}_w$, there is no
$SU(3)$ singlet which can be made with ${\cal D}_6$,  and no $SU(3)$
invariant
matrix element of the form (\ref{melm}) can be formed.  This is the well
known
result that $D^0-\D0bar$ mixing is prohibited by $SU(3)$ symmetry.

Now consider a single insertion of the $SU(3)$ violating spurion ${\cal
M}$.
The combination ${\cal D}_6{\cal M}$ transforms as $6\times
8=24+\15bar+6+\3bar$.  Note that there is still no invariant to be made
with
${\cal H}_w{\cal H}_w$.  It follows that $D^0-\D0bar$ mixing is not 
induced at first
order in
$SU(3)$ breaking.

With two insertions of ${\cal M}$, it becomes possible to make an $SU(3)$
invariant.  The decomposition of ${\cal D}{\cal M}{\cal M}$ is
\beqa
6\times(8\times 8)_S &=& 6\times(27+8+1)\nonumber\\
    &=& (60+\42bar+24+\15bar+\15bar'+6) + (24+\15bar+6+\3bar) + 6\,.
\eeqa
There are three elements of the $6\times 27$ part which can give
invariants
with  ${\cal H}_w{\cal H}_w$.  Each invariant yields a contribution
proportional to $s_1^2m_s^2$.  As promised, $D^0-\D0bar$
mixing arises only at second order in the $SU(3)$ violating parameter $m_s$.

\section{Estimating the size of $SU(3)$ breaking}\label{sec:estim}

We now turn to review some general estimates of the size of $SU(3)$
breaking
effects.  These effects can be approached from either an inclusive or an
exclusive point of view.  It is instructive to see how $SU(3)$ violation
appears in each case.

\subsection{``Inclusive" approach}

An elegant and concrete estimate of how $SU(3)$ violation enters $x$ and
$y$ is
the short distance analysis, first applied to $D^0-\D0bar$ mixing by
Georgi~\cite{Ge92} and later extended by other
authors~\cite{Oh93,Bigi:2000wn}.  We review it briefly, both to
establish the
contrast with our approach and to recall the results.  Let $\Lambda$ be 
a scale characteristic of the strong interactions, such as $m_\rho$ or 
$4\pi f_\pi$. In the limit $m_c
\gg\Lambda$, the momentum flowing through the light degrees of freedom 
in the
intermediate state is large and an operator production expansion (OPE)
can be
performed.  For example, one can write
\begin{equation}\label{gammaope}
\Gamma_{12} = \frac{1}{2 m_D}\, {\rm Im}\, \langle \D0bar |
    \,i\! \int\! {\rm d}^4 x\, T \Big\{
    {\cal H}^{\Delta C=1}_w (x)\, {\cal H}^{\Delta C=1}_w(0) \Big\}
    | D^0 \rangle \,,
\end{equation}
where ${\cal H}^{\Delta C=1}_w$ is the $|\Delta C|=1$ effective
Hamiltonian.  In
the OPE, the time ordered product in Eq.~(\ref{gammaope}) can be
expanded in
local operators of  increasing dimension; the higher dimension operators
are
suppressed by powers of $\Lambda/m_c$.

The leading contribution comes from the dimension-6 $|\Delta C|=2$
four-quark
operators corresponding to the short distance box diagram,
\begin{eqnarray}
O_1 &=& \bar u_\alpha \gamma_\mu P_L c_\alpha\,
    \bar u_\beta \gamma_\mu P_L c_\beta\,, \qquad
    O_1' = \bar u_\alpha P_L c_\alpha\,
    \bar u_\beta P_L c_\beta\,, \nonumber \\
O_2 &=& \bar u_\alpha \gamma_\mu P_L c_\beta\,
    \bar u_\beta \gamma_\mu P_L c_\alpha\,, \qquad
    O_2' = \bar u_\alpha P_L c_\beta\,
    \bar u_\beta P_L c_\alpha\,,
\end{eqnarray}
where $P_L = {1\over2}(1-\gamma_5)$.
If one neglects QCD running between $M_W$ and $m_c$, in which case $O_2$
and $O_2'$ do not contribute, one finds the simple expressions
\begin{eqnarray}
\Delta M_{\rm box} &=& {2 \over 3\pi^2}\, X_D\, {(m_s^2 - m_d^2)^2 \over
m_c^2}
    \left[ 1 - \frac54\, {B_D' \over B_D}\, {m_D^2\over (m_c + m_u)^2}
\right],
    \label{msd}\\
\Delta \Gamma_{\rm box} &=& {4 \over 3\pi}\, X_D\,
    {(m_s^2 - m_d^2)^2 \over m_c^2}\, {m_s^2 + m_d^2 \over m_c^2}\,
    \left[ 1 - \frac52\, {B_D' \over B_D}\, {m_D^2\over (m_c + m_u)^2}
\right].
    \label{gsd}
\end{eqnarray}
where $X_D=V_{cs}^2V_{cd}^2 G_F^2 m_D B_D f_D^2$, and $B_D^{(\prime)}$
are bag
factors for $O_1^{(\prime)}$, normalized to one in vacuum saturation.
Including leading logarithmic QCD effects enhances this estimate of
$\Delta\Gamma$ by approximately a factor of two~\cite{apreview}.
Eqs.~(\ref{msd}) and (\ref{gsd}) then lead to the estimates
\beq
x_{\rm box}\sim {\rm few}\times 10^{-5}\,,\qquad
    y_{\rm box}\sim {\rm few}\times 10^{-7}\,.
\eeq

Neglecting $m_d/m_s$, Eq.~(\ref{gsd}) is proportional to $m_s^6$. This
factor
comes from three sources:  (i) $m_s^2$ from an $SU(3)$ violating mass 
insertion
on
each quark line in the box graph; (ii) $m_s^2$ from an additional mass
insertion on each line to compensate the chirality flip from the first
insertion; (iii) $m_s^2$ to lift the helicity suppression for the decay
of a
scalar meson into a massless fermion pair.  The last factor of $m_s^2$ is
absent from Eq.~(\ref{msd}) for $\Delta M$; this is why at leading order
in the
OPE $y_{\rm box} \ll x_{\rm box}$.  Higher order terms in the OPE are
important, because the chiral suppressions can be lifted by quark
condensates
instead of by mass insertions, allowing $\Delta M$ and $\Delta\Gamma$ to
be
proportional to $m_s^2$.  This is the minimal suppression required by
$SU(3)$ symmetry, as we proved in
Sec.~\ref{sec:su3}.

\begin{table}
\begin{tabular}{|c|ccc|} \hline
ratio  &  4-quark  &  6-quark  &  8-quark \\ \hline\hline
~~$\Delta M/\Delta M_{\rm box}$~~  &  1  &  ~~$\Lambda^2 /m_s m_c$~~
    &  ~~$({\alpha_s/4\pi})(\Lambda^2/m_s m_c)^2$~~ \\ \hline
$\Delta \Gamma/\Delta M$  &  ~~$m_s^2/m_c^2$~~
    &  $\alpha_s/4\pi$  &  $\beta_0\,\alpha_s/4\pi $ \\ \hline
\end{tabular} \vspace{4pt}
\caption[]{The enhancement of $\Delta M$ and $\Delta\Gamma$ relative to
the
box diagram at various orders in the OPE.  $\Lambda$ denotes a hadronic
scale
around $4\pi f_\pi \sim 1\,$GeV.}
\label{opetable}
\end{table}

The order of magnitudes of the resulting contributions are summarized in
Table~\ref{opetable}.  In the first line, the contributions to $\Delta
M$ are
normalized to $\Delta M_{\rm box}$; in the second line, the
contributions to
$\Delta\Gamma$ are normalized to $\Delta M$ at each order.  The
contribution of
6-quark operators to $\Delta M$ is enhanced compared to the 4-quark
operators
by $\Lambda^2/m_c m_s$.  This can be as much as an order of magnitude, 
if we identify the hadronic scale $\Lambda$
as $4\pi f_\pi$~\cite{Manohar:1984md}.  The second chiral
suppression can
also be lifted, but only at the price of adding a hard gluon, so the
contribution of 8-quark operators to $\Delta M$ compared to the 6-quark
operators is $(\alpha_s/4\pi)(\Lambda^2/m_cm_s)$, which is of order
unity.\footnote{We disagree with Ref.~\cite{Bigi:2000wn}, in which it was
claimed
that $x$ and $y$ can arise at first order in $m_s$.  Such contributions
were
claimed to come from pseudogoldstone loops which diverge in the
infrared.  However,
there are no such divergences because the $\pi$, $K$ and $\eta$ are coupled
derivatively.  Such a contribution would also be in conflict with our
proof in Sec.~\ref{sec:su3} that $D$ mixing is second order in
$SU(3)$
violating effects.}  In the case of $\Delta\Gamma$, higher dimension
operators
are even more important~\cite{Bigi:2000wn}.  A 6-quark operator,
including a
hard gluon to give an on-shell intermediate state, lifts both a chiral
suppression and the helicity suppression.  The 8-quark operators
require a
second intermediate particle to contribute to $\Delta\Gamma$, which can
be
obtained by splitting the gluon already present for $\Delta M$ into a
quark
pair~\cite{Bigi:2000wn}, only costing a factor of $\beta_0\,
\alpha_s/(4\pi)
\sim 1$, where $\beta_0 = 11-{2\over3}n_f= 9$.  Thus, the dominant
contributions to $x$
are from 6- and 8-quark operators, while the dominant contribution to
$y$ is
from 8-quark operators.  With some assumptions about the hadronic matrix
elements, the resulting estimates are
\beq
    x \sim y \alt 10^{-3}\,.
\eeq
It is a general feature of OPE based analyses that $x\agt y$.  We 
emphasize that at this time these methods are useful for understanding 
the order of magnitude of $x$ and $y$, but not for obtaining reliable 
quantitative results. For example, to turn the estimates presented here 
into a systematic computation of $x$ and $y$ would require the 
calculation of almost two dozen nonperturbative matrix elements.

\subsection{``Exclusive" approach}

A long distance analysis of $D$ mixing is complementary to the OPE.
Instead of
assuming that the $D$ meson is heavy enough for duality to hold between
the
partonic rate and the sum over hadronic final states, here one assumes
that $D$
transitions are dominated by a small number of exclusive processes,
which are
examined explicitly.  This is particularly interesting for studying
$\Delta\Gamma$, which depends on real final states in $D$ decays.

For a long distance analysis, it is useful to express the width
difference
directly in terms of observable decay rates.  From Eq.~(\ref{ysum}), we
find
\beq\label{usey}
y = \sum_n \etackm(n)\, \etacp(n)\, \cos \delta_n\,
    \sqrt{\BR(D^0\to n)\, \BR(\D0bar\to n)} \,,
\eeq
where $\delta_n$ is the strong phase difference between the $D^0\to n$
and
$\D0bar\to n$ amplitudes.  In decays to many-body final states, the
strong
phases may have different values in different regions of the Dalitz
plot, in
which case the sum is supplemented by an integral over the Dalitz plot
for each
final state.  The CKM factor is $\etackm = (-1)^{n_s}$, where $n_s$ is
the
number of $s$ and $\bar s$ quarks in the final state.  For example,
$\etackm(K^+K^-) = +1$ and $\etackm(K^+ \pi^-) = -1$.    The factor
$\etacp = \pm
1$ is determined by the $CP$ transformation of the final state,
$CP|f\rangle =
\etacp|\bar f\rangle$, which is well-defined since $|f\rangle$ and $|\bar
f\rangle$ are in the same $SU(3)$ multiplet. This factor is the same for 
the whole multiplet.
For example,
$\eta_{CP}=+1$ for the decays to $K^+ K^-$, and therefore to all decays
into two pseudoscalars. For states where different
partial waves contribute with different $CP$ parities, $\etacp$ is
determined
separately for each partial wave. For example, $\etacp (\rho^+ \rho^-) =
+1$
for $\rho^+ \rho^-$ in a relative $s$ or $d$ wave, and $-1$ in a $p$
wave.
Finally, it is convenient to assemble the final states into $SU(3)$
multiplets
and write
\beq \label{useyy}
y = \sum_{a}\, y_{a} \,, \qquad
    y_a = \etacp(a) \sum_{n\in a} \etackm(n) \cos \delta_n\,
    \sqrt{\BR(D^0\to n)\, \BR(\D0bar\to n)} \,,
\eeq
where $a$ indexes complete $SU(3)$ multiplets.  By multiplets we refer
to the
$SU(3)$ representation of the entire final state, not of the individual 
mesons and baryons.

In practice, we cannot use Eq.~(\ref{useyy}) to get a reliable estimate 
of $y$,
since
the doubly Cabibbo suppressed rates have large errors, and there are
very little
data on strong phase differences.  To proceed
further, we would be forced to introduce model dependent assumptions
about the
amplitudes and/or their strong phases.  For example, in two-body $D$
decays to
charged pseudoscalars ($\pi^+\pi^-$, $\pi^+ K^-$, $K^+ \pi^-$, $K^+
K^-$), the
$SU(3)$ violation can enter through the decay rates or the strong phase
difference.  We know experimentally that in some of these rates the
$SU(3)$
breaking is sizable; for example $\BR(D^0 \to K^+ K^-)/ \BR(D^0 \to \pi^+
\pi^-) \simeq 2.8$~\cite{PDG}.  Such effects were the basis for the
claim in
Ref.~\cite{wolf} that $SU(3)$ is simply inapplicable to $D$ decays.  In
contrast, we know very little about the strong phase $\delta$ which
vanishes in
the $SU(3)$ limit; Ref.~\cite{Falk:1999ts} presented a model calculation
resulting in $\cos\delta \agt 0.8$, but it is also possible to obtain
much
larger values for $\delta$~\cite{cnp}.  Using Eq.~(\ref{useyy}), the
value of
$y_a$ corresponding to the $U$-spin doublet of charged $\pi$ and $K$ is
\begin{equation}
y_{\pi K} = \BR(D^0 \to \pi^+\pi^-) + \BR(D^0 \to K^+K^-)
    - 2\cos\delta\, \sqrt{\BR(D^0 \to K^-\pi^+)\, \BR(D^0 \to
K^+\pi^-)} \,.
\end{equation}
The experimental central values, allowing for $D$ mixing in the doubly
Cabibbo
suppressed rates, yield $y_{\pi K} \simeq (5.76 - 5.29 \cos\delta) \times
10^{-3}$~\cite{BGLNP}.  For small $\delta$ there is an almost perfect
cancellation even though the ratios of the individual rates significantly
violate $SU(3)$.  In the ``exclusive'' approach, $x$ is obtained from 
$y$ by use of a dispersion relation, and one generally finds $x\sim y$.

At this stage, one cannot use the exclusive approach to predict either 
$x$ or
$y$. Any estimate of their sizes depends on computing
$SU(3)$ breaking effects. While this problem is not tractable in
general, one source of $SU(3)$ breaking in $y$, from final state phase
space, can be calculated with only minimal and reasonable
assumptions. We will estimate these effects in the next
section.

\section{$SU(3)$ breaking from phase space}\label{sec:ps}

We now turn to the contributions to $y$ from on-shell final states.
There is a
contribution to the $D^0$ width difference from every common decay
product of
$D^0$ and $\D0bar$.  In the $SU(3)$ limit, these contributions cancel
when one
sums over complete $SU(3)$ multiplets in the final state.  The
cancellations
depend on $SU(3)$ symmetry both in the decay matrix elements and in the
final
state phase space.  While there are certainly $SU(3)$ violating
corrections to
both of these, it is extremely difficult to compute the $SU(3)$
violation in
the matrix elements in a model independent manner.\footnote{The $SU(3)$
breaking in matrix elements may be modest even in cases such as $D\to 
K^+ K^-$ and  $D\to \pi^+ \pi^-$, for which the ratio of measured rates 
appears
to be
very far from the $SU(3)$ limit~\cite{MJS}.}  However, with some mild
assumptions
about the momentum dependence of the matrix elements, the $SU(3)$
violation in
the phase space depends only on the final particle masses and can be
computed.
In this section we estimate the contributions to $y$ solely from $SU(3)$
violation in the phase space.\footnote{The phase space difference alone
can
explain the large $SU(3)$ breaking between the measured $D\to
K^*\ell\bar\nu$
and $D\to \rho\ell\bar\nu$ rates, assuming no $SU(3)$ breaking in the
form
factors~\cite{LSW}.  Recently it was shown
that the
lifetime ratio of the $D_s$ and $D^0$ mesons may also be explained this
way~\cite{NP}.}  We will find that this source of $SU(3)$ violation can
generate $y$ of the order of a percent.

The mixing parameter $y$ may be written in terms of the matrix elements
for
common final states for $D^0$ and $\D0bar$ decays,
\beq
y = {1\over\Gamma} \sum_n \int [{\rm P.S.}]_n\,
    \langle \D0bar|\,{\cal H}_w\,|n \rangle \langle n|\,{\cal H}_w\,|D^0 
\rangle\,,
\eeq
where the sum is over distinct final states $n$ and the integral is over
the
phase space for state $n$.  Let us now perform the phase space integrals
and
restrict the sum to final states $F$ which  transform within a
single $SU(3)$ multiplet $R$.  The result is a  contribution to $y$ of
the form
\beq
    {1\over\Gamma}\, \langle\D0bar|\,{\cal H}_w
     \bigg\{ \eta_{CP}(F_R)\sum_{n\in  F_R}
    |n\rangle \rho_n\langle n| \bigg\} {\cal H}_w\,|D^0\rangle\,,
\eeq
where $\rho_n$ is the phase space available to the state $n$.  In the
$SU(3)$
limit, all the $\rho_n$ are the same for $n\in F_R$, and the quantity in
braces
above is an $SU(3)$ singlet.  Since the $\rho_n$ depend only on the known
masses of the particles in the state $n$, incorporating the true values
of
$\rho_n$ in the sum is a calculable source of $SU(3)$ breaking.

This method does not lead directly to a calculable contribution to $y$,
because
the matrix elements $\langle n|{\cal H}_w|D^0\rangle$ and
$\langle\D0bar|{\cal H}_w|n\rangle$
are not known.  However, $CP$ symmetry, which in the Standard Model and
almost all scenarios of new physics is to an excellent approximation
conserved in $D$ decays, relates $\langle\D0bar|{\cal H}_w|n\rangle$ to 
$\langle
D^0|{\cal H}_w|\overline{n}\rangle$. Since $|n\rangle$ and
$|\overline{n}\rangle$ are
in a common $SU(3)$ multiplet,  they are determined by a single effective
Hamiltonian. Hence the ratio
\beq\label{yfr}
   y_{F,R} = {\sum_{n\in F_R} \langle\D0bar|\,{\cal H}_w|n\rangle \rho_n
    \langle n|{\cal H}_w\,|D^0\rangle \over
    \sum_{n\in F_R} \langle D^0|\,{\cal H}_w |n\rangle \rho_n
    \langle n|{\cal H}_w\,|D^0\rangle}
   = {\sum_{n\in F_R} \langle\D0bar|\,{\cal H}_w|n\rangle \rho_n
    \langle n|{\cal H}_w\,|D^0\rangle \over \sum_{n\in F_R}
   \Gamma(D^0\to n)}
\eeq
is calculable, and represents the value which $y$ would take if elements
of
$F_R$ were the only channel open for $D^0$ decay.  To get a true
contribution
to $y$, one must scale $y_{F,R}$ to the total branching  ratio to all the
states in $F_R$.  This is not trivial, since a given physical final state
typically decomposes into a sum over more than one multiplet $F_R$.  The
numerator of $y_{F,R}$ is of order $s_1^2$ while the  denominator is of
order
1, so with large $SU(3)$ breaking in the phase space the natural size of
$y_{F,R}$ is 5\%.

In this analysis, phase space is the only source of $SU(3)$ violation
which we
will include. Of course, there are other $SU(3)$ violating effects, such
as in
matrix elements and final state interaction phases.  The purpose of our
calculation is to explore the rough size of
$SU(3)$ violation in exclusive contributions to $y$.  We assume that
there is no cancellation with other sources of $SU(3)$ breaking, or
between the various multiplets which occur in $D$ decay, that would
reduce our result for $y$ by an order of magnitude.
  This is equivalent to assuming that the $D$ meson
is not
heavy enough that duality can be expected to enforce such
cancellations.

We begin by computing $y_{F,R}$ for $D$ decays to states $F=PP$
consisting of
a pair of pseudoscalar mesons  such as $\pi$, $K$,  $\eta$.  We neglect $\eta-\eta'$ mixing throughout this analysis, and we have checked that this simplification has a negligible effect on the numerical results. Since $PP$ 
is
symmetric in the two mesons, it must transform as an element of $(8\times
8)_S=27+8+1$.  In principle, there are three possible amplitudes for
$D^0\to
PP$, one with the pair in a 27 and ${\cal H}_w$ in a $\15bar$,
\beq\label{27inv}
    A_{27}(PP_{27})^{km}_{ij}H^{ij}_kD_m\,,
\eeq
one with the pair in an 8 and and ${\cal H}_w$ in a $\15bar$,
\beq\label{8inv1}
    A_8^{\15bar}(PP_8)^k_iH^{ij}_kD_j\,,
\eeq
and one with the pair in an 8 and and ${\cal H}_w$ in a $6$,
\beq\label{8inv2}
    A_8^6(PP_8)^k_iH^{ij}_kD_j\,.
\eeq
However, the product $H^{ij}_kD_j$ with $(ij)$ symmetric (the $\15bar$)
is
proportional to $H^{ij}_kD_j$ with $(ij)$ antisymmetric (the 6), and the
linear
combination $A_8\equiv A_8^{\15bar}-A_8^6$ is the only one which
appears.  Thus
there are effectively two invariant amplitudes.  There is no $SU(3)$ 
invariant amplitude to produce the final state in an  singlet.  Note 
that since we are assuming $SU(3)$ symmetry in the matrix elements, such 
final states do not appear in our analysis.

It is straightforward to use these invariants in Eq.~(\ref{yfr}) to 
compute $y_{F,R}$.  As an example, for  $y_{PP,8}$ we obtain
\beqa
y_{PP,8} &=& s_1^2\,\bigg[ \frac12\, \Phi(\eta,\eta)
   + \frac12\,  \Phi(\pi^0,\pi^0) + \frac13\, \Phi(\eta,\pi^0)
   + \Phi(\pi^+,\pi^-) + \Phi(K^+,K^-) - \frac16\,
\Phi(\eta,K^0) \nonumber\\*
&&\quad{}  - \frac16\, \Phi(\eta,\K0bar) - \Phi(K^+,\pi^-)
    - \Phi(K^-,\pi^+) - \frac12\, \Phi(K^0,\pi^0)
    - \frac12\, \Phi(\K0bar,\pi^0) \bigg] \nonumber\\*
&&{} \times \bigg[ \frac16\, \Phi(\eta,\K0bar) + \Phi(K^-,\pi^+)
   + \frac12\, \Phi(\K0bar,\pi^0) + {\cal O}(s_1^{2}) \bigg]^{-1} \,,
\eeqa
where $\Phi(P_1,P_2)$ is the phase space integral for the decay into
mesons $P_1$ and $P_2$.
In a two-body decay, $\Phi(P_1,P_2)$ is proportional to $|\vec
p\,|^{2\ell+1}$, where $\vec p$ and $\ell$ are the spatial momentum and
orbital angular momentum of the final state particles.
   For $D^0\to PP$, the decay is into an $s$ wave.  It is
straightforward to compute the required ratios from the known
pseudoscalar
masses,
\beq\label{mmratios}
    y_{PP,8} = -0.0038\, s_1^2 = -1.8\times 10^{-4}\,, \qquad
    y_{PP,27} = -0.00071\, s_1^2 = -3.4\times 10^{-5}\,.
\eeq
These effects are no larger than one finds in the inclusive analysis.
This is
not surprising, since as in the parton picture, the final states  are
far from
threshold.

Next we turn to final states of the form $PV$, consisting of a
pseudoscalar and
a vector meson.  Note that three-body final states $3P$ can resonate
through
$PV$, and so are partially included here.  In this case there is no
symmetry
between the mesons, so in principle all representations in the
combination
$8\times 8=27+10+\tenbar+8_S+8_A+1$ can appear.  For simplicity, we
take the quark content of the $\phi$ and $\omega$ respectively to be $\bar ss$ and $(\bar uu+\bar dd)/\sqrt2$, and consider only the combination which appears in the $SU(3)$ octet.  We have checked that reasonable variations of the $\phi -\omega$ mixing angle have a negligible effect on our numerical results. For each
representation, there is a single invariant, up to the same degeneracy
for the 8 as in the $PP$ case.  Along with the analogues of
Eqs.~(\ref{27inv})--(\ref{8inv2}) with coefficients $B_{27}$ and
$B_8\equiv
B_8^{\15bar}-B_8^6$, we have the new invariants
\beq
    B_{10}(PV_{10})_{ijk}H^{ij}_mD_n\epsilon^{kmn}
\eeq
for ${\cal H}_w$ in a $\15bar$, and
\beq
    B_{\tenbar}(PV_{\tenbar})^{ijk}H^{lm}_iD_j\epsilon_{klm}
\eeq
for ${\cal H}_w$ in a 6.  It turns out that these two invariants are
proportional to
each other.  As before, the $SU(3)$ singlet final state is not produced.

Both because one of the particles is more massive, and
because the
decay is now into a $p$ wave, the phase space dependence is stronger
than for
the $PP$ final state.  We obtain the ratios
\beqa\label{mvratios}
&& y_{PV,8_S} = 0.031\, s_1^2 =0.15\times 10^{-2}\,,\qquad
    y_{PV,8_A} =0.032\, s_1^2 = 0.15\times 10^{-2}\,,\nonumber\\
&& y_{PV,10} =0.020\, s_1^2 = 0.10\times 10^{-2}\,,\qquad\,
    y_{PV,\tenbar} = 0.016\, s_1^2 = 0.08\times 10^{-2}\,,\nonumber\\
&& y_{PV,27} = 0.040\, s_1^2 = 0.19\times 10^{-2}\,.
\eeqa
For any representation of the final state, 
the effects are less than one percent.

For the $VV$ final state, decays into $s$, $p$ and $d$ waves are all
possible.
Bose symmetry and the restriction to zero total angular momentum
together imply
that only the symmetric $SU(3)$ combinations appear.
Because some $VV$ final states, such as $\phi K^*$, lie near the $D$ threshold, the inclusion of vector meson widths is quite important.  Our model for the resonance line shape is a Lorentz invariant Breit-Wigner normalized on $0\le m<\infty$, 
\beq
  f(m;m_R,\Gamma_R)=N(m_R,\Gamma_R)\,
  {m^2\,\Gamma_R^2\over (m^2-m_R^2)^2+m^2\Gamma_R^2}\,,
\eeq
where $m_R$ and $\Gamma_R$ are the mass and width of the vector meson, and $m^2$ is the square of its four-momentum in the decay.
  For $s$ wave
decays, we
find the ratios
\beq
    y_{VV,8} = -0.081\, s_1^2 = -0.39\times 10^{-2}\,,\qquad
    y_{VV,27} = -0.061\, s_1^2 = -0.30\times 10^{-2}\,,
\eeq
while for $p$ wave decays we find
\beq
    y_{VV,8} = -0.10\, s_1^2 = -0.48\times 10^{-2}\,,\qquad
    y_{VV,27} = -0.14\, s_1^2 = -0.70\times 10^{-2}\,,
\eeq
and for $d$ waves,
\beq
    y_{VV,8} = 0.51\, s_1^2 = 2.5\times 10^{-2}\,,\qquad
    y_{VV,27} = 0.57\, s_1^2 = 2.8\times 10^{-2}\,.
\eeq
With these heavier final states and with the higher partial waves, we
see that
effects at the level of a percent are quite generic.  The vector meson widths turn out to be quite important; if they were neglected, the results in the $p$- and $d$-wave channels would be larger by approximately a factor of three.  The finite widths soften the $SU(3)$ breaking which otherwise would be induced by a sharp phase space boundary.  We have checked that our results are not very sensitive to variations in
the line shape used to model the vector meson widths.  Again, $4P$ and
$PPV$ final states can  resonate through $VV$, so they are partially
included
here.  Our results for two-body final states are  summarized in
Table~\ref{ytwobody}.

\begin{table}
\begin{tabular}{|@{~~~}lc|c|c|} \hline
\multicolumn{2}{|c|}{~~Final state representation~~~}  &
     ~~~$y_{F,R}/s_1^2$~~~ & ~~~$y_{F,R}\ (\%)$~~~  \\ \hline\hline
    $PP$  &  $8$  &  $-0.0038$ & $-0.018$  \\
    &  $27$  &  $-0.00071$  & $-0.0034$ \\ \hline
    $PV$  &  $8_S$  &  $0.031$ & $0.15$\\
    &  $8_A$  &  $0.032$  & $0.15$ \\
    &  $10$  &  $0.020$ & $0.10$ \\
    &  $\overline{10}$  &  $0.016$ & $0.08$ \\
    &  $27$  &  $0.040$  & $0.19$\\ \hline
    $(VV)_{\mbox{$s$-wave}}$  &  $8$  &  $-0.081$ & $-0.39$ \\
    &  $27$  &  $-0.061$ & $-0.30$\\
    $(VV)_{\mbox{$p$-wave}}$  &  $8$  &  $-0.10$ & $-0.48$\\
    &  $27$  &  $-0.14$ & $-0.70$ \\
    $(VV)_{\mbox{$d$-wave}}$  &  $8$  &  $0.51$ & $2.5$ \\
    &  $27$  &  $0.57$  & $2.8$\\ \hline
\end{tabular} \vspace{4pt}
\caption{Values of $y_{F,R}$ for two-body final states.  This represents 
the value which $y$ would take if elements of $F_R$ were the only 
channel open for $D^0$ decay.}
\label{ytwobody}
\end{table}

As we go to final states with more particles, the combinatoric
possibilities
begin to proliferate.  We will consider the final states $3P$ and $4P$,
and
for concreteness require that the pseudoscalars be found in a totally
symmetric
8 or 27 representation of $SU(3)$. This assumption is convenient,
because the
phase space integration is much simpler if it can be performed
symmetrically.
These final states should be representative; we have no reason to
believe that
this choice selects final state multiplets for which phase space effects
are
particularly enhanced or suppressed.  Note that $3\times(\15bar+6)$
contains
no representation larger than a 27.

In contrast to the two-body case, for three-body final states the 
momentum dependence of the matrix
elements is no longer fixed by the conservation of angular momentum.  
The simplest assumption is to take a momentum
independent matrix element, with all three final state particles in an
$s$
wave.  In that case, we find
\beq
    y_{3P,8} = -0.48\, s_1^2 = -2.3\times 10^{-2}\,,\qquad
    y_{3P,27} = -0.11\, s_1^2 =- 0.54\times10^{-2}\,.
\eeq
Note that the $SU(3)$ violation is smaller for the larger multiplets, as
more
final states enter the sum.  It may be that the 8 is in some sense an
unusually
small representation for three or more particles, and that this mode
enhances the
$SU(3)$ violation by providing fewer distinct final states among which
cancellations can occur.  The enhancement of $y_{3P,8}$ over $y_{3P,27}$
is
not a peculiarity of $s$ wave decays.  We have also considered other 
matrix elements;
for example, if one of the mesons has angular momentum $\ell = 1$ in the 
$D^0$ rest frame
(balanced by the combination of the other two mesons), then the ratios 
become
\beq
    y_{3P,8} = -1.13\, s_1^2 = -5.5\times 10^{-2}\,,\qquad
    y_{3P,27} = -0.074\, s_1^2 = -0.36\times 10^{-2}\,.
\eeq
Alternatively, we could imagine introducing a mild ``form factor
suppression,''
with a weight such as $\Pi_{i\ne j}(1 - m_{ij}^2/Q^2)^{-1}$, where 
$m_{ij}^2=(p_i+p_j)^2$, and
$Q=2\,{\rm
GeV}$ is a typical resonance mass.  The result then changes to
\beq
    y_{3P,8} =- 0.44\, s_1^2 =- 2.1\times 10^{-2}\,,\qquad
    y_{3P,27} = -0.13\, s_1^2 = -0.64\times10^{-2}\,.
\eeq

Finally, we have studied the final state with four pseudoscalars, with
the
mesons in an overall symmetric 8 or a symmetric 27.  We take a momentum
independent matrix element.  There are actually two symmetric 27
representations; we call the 27 the representation of the form
$R^{ij}_{kl}=[M^i_mM^m_kM^j_nM^n_l+{\rm symmetric}-{\rm traces}]$ and
the $27'$
the one of the form $R^{ij}_{kl}=[M^i_mM^m_nM^n_kM^j_l+{\rm
symmetric}-{\rm
traces}]$.  Then we find
\beqa
&& y_{4P,8} = 3.3\, s_1^2 = 16\times 10^{-2}\,,\qquad\,
    y_{4P,27} = 2.2\, s_1^2 = 11\times 10^{-2}\,.\nonumber\\
&& y_{4P,27'} = 1.9\, s_1^2 = 9.2\times 10^{-2}\,.
\eeqa
Here the partial contributions to $y$ are very large, of the order of
10\%.
This is not surprising, since $4P$ final states containing more than one
strange particle are close to $D$ threshold, and the ones containing no
pions
are kinematically inaccessible.  There is no reason to expect $SU(3)$
cancellations to persist effectively in this regime.  Our results for
$3P$
and $4P$ final states are summarized in Table~\ref{mltybodytbl}.

\begin{table}
\begin{tabular}{|@{~~~}lc|c|c|} \hline
\multicolumn{2}{|c|}{~~Final state representation~~~}  &
    ~~~$y_{F,R}\,/s_1^2$~~~ & ~~~$y_{F,R}\ (\%)$~~~\\ \hline\hline
$(3P)_{\mbox{$s$-wave}}$	&  $8$  &  $-0.48$  & $-2.3$\\
    &  $27$  &  $-0.11$  & $-0.54$ \\
$(3P)_{\mbox{$p$-wave}}$	&  $8$  &  $-1.13$  & $-5.5$ \\
    &  $27$  &  $-0.07$   & $-0.36$  \\
$(3P)_{\mbox{form-factor}}$	&  $8$  &  $-0.44$  & $-2.1$\\
    &  $27$  &  $-0.13$ & $-0.64$ \\ \hline
$4P$  &  $8$  &  $3.3$ & $16$  \\
    &  $27$  &  $2.2$  & $9.2$ \\
    &  $27'$  &  $1.9$ & $11$ \\ \hline
\end{tabular} \vspace{4pt}
\caption[]{Values of $y_{F,R}$ for three- and four-body final states.}
\label{mltybodytbl}
\end{table}

Formally, one could construct $y$ from the individual $y_{F,R}$ by
weighting
them by their $D^0$ branching ratios,
\beq\label{ycombine}
    y = {1\over\Gamma} \sum_{F,R}\, y_{F,R}
    \bigg[\sum_{n\in F_R}\Gamma(D^0\to n)\bigg]\,.
\eeq
However, the data on $D$ decays are neither abundant nor precise enough
to
disentangle the decays to the various $SU(3)$ multiplets, especially
for the
three- and four-body final states.  Nor have we computed  $y_{F,R}$ for
all or
even most of the available representations.   Instead, we can only
estimate
individual contributions to $y$ by  assuming that the representations
for which
we know $y_{F,R}$ to be  typical for final states with a given
multiplicity,
and then to scale to  the total branching ratio to those final states.
The
total branching  ratios of $D^0$ to two-, three- and four-body final
states can
be extracted from Ref.~\cite{PDG}.  The results are presented in
Table~\ref{branching}, where we round to the nearest 5\% to emphasize the
uncertainties in these numbers.  Close to half of all $D^0$ decays are
accounted for in this table; the rest are decays to other modes such as 
$PPV$, decays to states with $SU(3)$ singlet mesons,
decays to
higher resonances, semileptonic decays, and other suppressed
processes.
Based on data in the channel ${\overline K}{}^{0*}\rho^0$, the $VV$ 
final state is
dominantly $CP$ even, consistent with an equal distribution between $s$
and
$d$ wave decays (although favoring a small $s$ wave enhancement).

\begin{table}
\begin{tabular}{|c|c|} \hline
Final state			&  ~~fraction~~  \\ \hline\hline
    $PP$				&  5\%  \\ \hline
    $PV$				&  10\%  \\ \hline
    ~~$(VV)_{\mbox{$s$-wave}}$~~	&  5\%  \\
    $(VV)_{\mbox{$d$-wave}}$	&  5\%  \\ \hline
    $3P$				&  5\%  \\ \hline
    $4P$			&  10\%  \\ \hline
\end{tabular} \vspace{4pt}
\caption[]{Total $D^0$ branching fractions to classes of final states,
rounded
to nearest 5\%~\cite{PDG}.}
\label{branching}
\end{table}

We estimate the contribution  to $y$ from a given type of final state by taking the product of the typical $y_{F,R}$ found in our calculation with the approximate branching ratios given in Table~\ref{branching}.  Such estimates are necessarily crude, but they are sufficient to give a sense of the order of magnitude of $y$ which is to be expected.  While in most cases the contributions are small, of the order of $10^{-3}$ or less, we observe that 
$D^0$ decays to nonresonant $4P$ states naturally contribute to $y$ at the percent level.  The reason for such unusually large $SU(3)$ violating effects in $y$ is that approximately $10\%$ of $D^0$ decays are to final states for which the complete $SU(3)$ multiplets are not kinematically accessible.

It should be noted that for $D$ decays to final states so close to threshold, our argument that $D$ mixing is second order in $SU(3)$ violation is inapplicable, because its underlying assumption that $SU(3)$ violation enters perturbatively is not met.  In particular the proof fails near $D$ threshold, if the decay is either to weakly decaying final states or to hadrons with widths $\Gamma$ which are smaller than $m_s$.  In either case, the phase space available for the decay can vary rapidly on the scale of $m_s$, spoiling the analytic expansion.  For decays to hadronic resonances, $\Gamma/m_s$  is a small parameter which is not analytic as $m_s\to0$.  For decays to long lived mesons, the $\Theta$-functions which fix the boundaries of phase space are not analytic functions of their arguments, which in turn depend on $m_s$ through the masses of the final state hadrons.  In this way, the generic $m_s^2/m_D^2$ suppression is lifted and we find larger $SU(3)$ violation in $y$ just at the point that the conditions of the proof are not satisfied.  We will see a similar failure of $SU(3)$ cancellations when we study $D$ mixing induced by resonances in Sec.~\ref{sec:res}.

We have not considered all possible final states which might give large contributions to $y$.  In particular, the branching ratio for $D^0\to K^-a_1^+$ is $(7.3\pm1.1)\%$~\cite{PDG}, even though this final state is quite close to $D$ threshold.  Unfortunately, the identities of the $SU(3)$ partners of the $a_1(1260)$, which has $J^{PC}=1^{++}$, are not well established.  While it is natural to identify the $K_1(1400)$ as the corresponding strange axial vector meson, and the $f_1(1285)$ as the analogue of the $\omega$, there is no natural candidate for the $\bar ss$ analogue of the $\phi$.  The size of $y_{PV^*}$ is quite sensitive to this choice, as well as to the value taken for the poorly measured width of the $a_1$. If we take the $\bar ss$ state to be the $f_1(1420)$, and $\Gamma(a_1)=400\,$MeV, we find $y_{PV^*,8_S}=1.8\%$.  If instead we take the $f_1(1510)$, we find $y_{PV^*,8_S}=1.7\%$.  With $\Gamma(a_1)=250\,$MeV, these numbers become $2.5\%$ and $2.4\%$, respectively.  Although it is clear that percent level contributions to $y$ are possible from $SU(3)$ violation in this channel, the data are still too poor to draw firm conclusions.

On the basis of this analysis, in particular as applied to the $4P$ final state, we would conclude that $y$  on the order  of a
percent would be completely natural.  Anything an order of
magnitude smaller would require significant  cancellations which do not
appear
naturally in this framework.  Cancellations would be expected only if 
they were enforced by the OPE, that is, if the charm quark were heavy 
enough that the ``inclusive'' approach were applicable.  The hypothesis 
underlying the present analysis is that this is not the case.

\section{$SU(3)$ breaking from nearby resonances}\label{sec:res}

One interesting feature of the $D^0$ is that there are excited
mesons
with masses close to $m_D$.  As a result, it would not be unnatural for
$K$
resonances to play an important role in $D$ decays.  This possibility has
already been explored in the
literature~\cite{Golowich:1998pz,Falk:1999ts,Golowich:1981yg,Gronau:1999zt}
.  Here we explore $SU(3)$ breaking in the resonance contribution
to $D$ mixing.

We are interested in the process $D^0\to R\to \D0bar$, where $R$ is a
resonance with mass $m_R$ and width $\Gamma_R$.  Only spin zero
resonances are
relevant.  The contribution of a single state to the $D$  mass and width
differences is given by
\begin{equation}\label{xyres}
y^{\rm res}_R= \eta_R\,\frac{|H_R|^2}{\Gamma}\,
    {\Gamma_R \over (m_D^2 - m_R^2)^2 + m_D^2 \Gamma_R^2} \,, \qquad
x^{\rm res}_R= \eta_R\,\frac{2 |H_R|^2}{\Gamma\, m_D}\,
    {m_D^2 - m_R^2 \over (m_D^2 - m_R^2)^2 + m_D^2 \Gamma_R^2} \,.
\label{delmg}
\end{equation}
where $|H_R|^2 \equiv \langle \D0bar | {\cal H}_W | R \rangle \langle
R | {\cal
H}_W | D^0 \rangle$ parameterizes the couplings of $R$ to $D^0$ and
$\D0bar$, and $\eta_R$ is the $CP$ eigenvalue of the  $SU(3)$ multiplet 
within which the resonance resides.
If we assume the absence  of direct $CP$ violation in $D$ decays, then
$\langle
\D0bar | {\cal  H}_W | R \rangle$ may be related to $\langle R | {\cal
H}_W |
D^0 \rangle$ by $SU(3)$.  The ratio
\beq
    {x^{\rm res}_R \over y^{\rm res}_R} = {2(m_D^2 - m_R^2) \over m_D\,
\Gamma_R} \,,
\eeq
is independent of $H_R$.  Significant contributions to $x$ and $y$ from
the
resonance $R$ are possible only if $m_D^2 - m_R^2 \alt m_D \Gamma_R$.

As a concrete example, consider the $K^*(1950)$, a positive parity 
excited kaon which, because of its large width,
may play an important role in mediating $D^0\to K^-\pi^+$.  Fitting the
$K^*(1950)$ contribution to the observed $D\to K\pi$ rates, one finds
that
resonance mediation could be as large as the usual quark tree
amplitude~\cite{Falk:1999ts}.  We can estimate an upper bound on the
contribution of $K^*(1950)$ to $y$ by assuming that the resonance is
completely
responsible for $D\to K\pi$.  The limit is given by
\beq
{|\Delta\Gamma|\over\Gamma}\le { \langle \D0bar | {\cal H}_W | K_H 
\rangle
    \langle K_H | {\cal H}_W | D^0 \rangle \over
    \langle D^0 | {\cal H}_W | K_H \rangle \langle K_H | {\cal H}_W | D^0
\rangle}
    \times {\BR(D^0\to K\pi)\over \BR(K_H\to K\pi)}\,,
\eeq
where we denote the $K^*(1950)$ by $K_H$.  With $\BR(D^0\to K\pi) \simeq
6\%$ and $\BR(K_H\to K\pi) \simeq52\%$, we find  $|y| \leq 0.06\, s_1^2
\simeq 3\times10^{-3}$.  If $D$ mixing is mediated by a resonance, then 
we expect $x$ and $y$ to be roughly of the same size.

This upper bound is too generous, because we have not included the
suppression from $SU(3)$ cancellations.  Note that our proof of 
Sec.~\ref{sec:su3}, that $SU(3)$ violation appears only at second order 
in $m_s$, applies only so long as $m_s\ll\Gamma_R$.  While this must be 
true in the limit $m_R\sim m_D\to\infty$, in which case $\Gamma_R$ 
scales as $m_c$, the ratio $m_s/\Gamma_R$ may not be small for 
resonances near the physical $D$ mass.  Therefore, $SU(3)$ cancellations 
may be less effective for resonances than for real final states.

The resonances in question fall into a positive parity 8 of $SU(3)$, consisting of 
states which we will denote
$(\pi_H,K_H,\eta_H)$.  If these states were degenerate and had equal
widths,
their contributions to $D$ mixing would cancel.  A measure of the actual
effectiveness of this cancellation is the contribution of the entire
multiplet
relative to that of the $K_H$.  The $SU(3)$ partners of the $K^*(1950)$
have
not been conclusively identified.  Instead of speculating,
we will explore the efficiency of $SU(3)$ cancellations qualitatively by
taking
the simple mass relations
\beq
    m_{\pi_H} = m_{K_H} - m_s\,, \qquad
    m_{\eta_H} = m_{K_H} + {1\over3}\, m_s\,,
\eeq
and assuming that the widths of the $\pi_H$ and $\eta_H$ are the same as
$\Gamma(K_H)\simeq200\,$MeV.  Then
\beq
y^{\rm res} = y^{\rm res}_{K_H} -{1\over4}\, y^{\rm res}_{\pi_H}
    -{3\over4}\, y^{\rm res}_{\eta_H}\,.
\eeq
For $m_s=150\,$MeV, we find $y^{\rm res}/y^{\rm res}_{K_H}=0.27$.  The
cancellations are somewhat less effective for $x^{\rm res}$, with
$x^{\rm res}/x^{\rm res}_{K_H}=0.50$. We see that even for the 
$K^*(1950)$, likely to
be the
most favorable for inducing a large effect,  $SU(3)$ cancellations
reduce the
contributions to $x^{\rm res}$ and $y^{\rm res}$.
We conclude that it would be quite unlikely for resonances to make a
contribution to $y$ at the level  of one percent.

\section{Conclusions}\label{sec:concl}

The motivation most often cited in searches for $D^0 - \D0bar$ mixing is 
the
possibility of observing a signal from new physics which may dominate 
over the
Standard Model contribution.  But to look for new physics in this way, 
one must
be confident that the Standard Model prediction does not already 
saturate the
experimental bound.  Previous analyses based on short distance 
expansions have
consistently found $x,y\alt10^{-3}$, while na\"\i ve estimates based on 
known
$SU(3)$ breaking in charm decays allow an effect an order of magnitude 
larger.
Since current experimental sensitivity is at the level of a few percent, 
the
difference is quite important.

In this paper we have performed a general $SU(3)$ analysis of the 
contributions to
$y$.  We proved that if $SU(3)$ violation may be treated perturbatively, 
then $D^0 - \D0bar$ mixing in the Standard Model is generated only at 
second order in $SU(3)$ breaking effects.
Within the exclusive approach, we identified an $SU(3)$ breaking effect,
$SU(3)$ violation in final state phase space, which can be calculated
with minimal model dependence.  We found that phase space effects
alone can provide enough $SU(3)$ breaking to induce $y\sim10^{-2}$.  Large effects in $y$ appear for decays to final states close to $D$ threshold, where an analytic expansion in $SU(3)$ violation is no longer possible.

We believe
that this is an important result.  Despite the large 
uncertainties, this
is the first model independent calculation to give $y$ close to the 
present
experimental bounds. While some degree of cancellation is possible between 
different
multiplets, as would be expected in the $m_c\to \infty$ limit, or between
$SU(3)$ breaking in phase space and in matrix elements, it is not known 
how effective these cancellations are.  The most
reasonable assumption in light of our analysis is that they are not
significant enough to result in an order of magnitude suppression of
$y$.
Therefore, any future discovery of a $D$ meson width difference should 
not by itself be
interpreted as an indication of the breakdown of the Standard Model.

However, our analysis does not amount to a Standard Model calculation of 
$y$. First, we have considered only $SU(3)$ breaking from phase
space, and have not included any symmetry breaking in the matrix 
elements.
Second, we have not calculated the contributions from all final states.  
Had we
done so, we would still need very precise experimental data in order to
disentangle the various $SU(3)$ multiplets  and combine the results into 
an
overall value of $y$.  Third, we have assumed that the charm quark is 
not heavy
enough for duality to enforce significant cancellations between the 
various
nonleptonic $D$ decay channels, although some degree of cancellation is 
to be
expected.

The implication of our results for the Standard Model prediction for  
$x$ is
less apparent. While analyses based on the ``inclusive" approach 
generally
yield  $x \agt y$, it is not clear what the ``exclusive" approach 
predicts.
The effect of $SU(3)$ breaking in phase space in $x$ is softer than in 
$y$, so
one would expect $x<y$ from our analysis.  Thus if $x > y$ is found
experimentally, it may still be an indication of a new physics 
contribution to
$x$, even if $y$ is also large.  On the other hand, if $y>x$ then it 
will be
hard to find signals of new physics, even if such contributions dominate 
$\Delta M$.  The linear sensitivity to new physics in the analysis of the
time dependence of $D^0\to K^+\pi^-$ is from $x' = x \cos\delta + y
\sin\delta$ and $y'= y \cos\delta - x \sin\delta$ instead of $x$ and 
$y$.  If
$y> x$, then $\delta$ would have to be known precisely for these terms 
to be
sensitive to new physics in $x$.

There remain large uncertainties in the Standard Model predictions of 
$x$ and
$y$, and values near the current experimental bounds cannot be ruled out.
Therefore, it will be difficult to find a clear indication of physics 
beyond the
Standard Model in $D^0 - \D0bar$ mixing measurements.  We believe that 
at this stage the only robust potential signal of new
physics in $D^0 - \D0bar$ mixing is $CP$ violation, for which the
Standard Model prediction is very small. Unfortunately, if $y$ is larger
or much larger than $x$, then the observable $CP$ violation in $D^0 - 
\D0bar$
mixing is necessarily small, even if new physics dominates $x$.  
Therefore, searching for new physics and $CP$ violation in $D^0-\D0bar$ 
mixing should aim at precise measurements of both $x$ and $y$, and at 
more complicated analyses involving the extraction of the strong
phase in the time dependence of doubly  Cabibbo suppressed decays.

\acknowledgments

It is a pleasure to acknowledge helpful discussions with Yossi Nir, 
Helen Quinn and Martin Savage.
We thank the Aspen Center for Physics for hospitality while portions of
this work were completed.
A.F.~was  supported in part by the U.S.~National Science Foundation
under Grant
PHY--9970781, and is a Cottrell Scholar of
the
Research  Corporation. Y.G.~was supported in part by
the Israel Science Foundation under Grant No.~237/01-1, and by the 
Technion
V.P.R Fund -- Harry Werksman Research Fund.
Z.L.~was supported in part by the Director, Office of  Science,  Office
of High
Energy and Nuclear Physics, Division of High Energy  Physics,  of the
U.S.\
Department of Energy under Contract DE-AC03-76SF00098.
The work of
Y.G.\ and Z.L.\ was also supported in part by the United
States--Israel Binational Science Foundation (BSF) through Grant No.
2000133.  A.P.~thanks the Cornell University Theory Group, where part of 
this work was performed.


\begin{thebibliography}{99}


\bibitem{Ge92}
H. Georgi, Phys. Lett. B297, 353 (1992).

\bibitem{Oh93}
T. Ohl, G. Ricciardi and E. H. Simmons, Nucl. Phys. B403, 605 (1993).

\bibitem{ap}
A.A. Petrov, Phys. Rev. D56, 1685 (1997).

\bibitem{Da85}
A. Datta and M. Khambakhar, Zeit. Phys. C27, 515 (1985).

\bibitem{dght}
J. Donoghue, E. Golowich, B. Holstein and J. Trampetic,
Phys. Rev. D33, 179 (1986).

\bibitem{BGLNP}
S.~Bergmann, Y.~Grossman, Z.~Ligeti, Y.~Nir and A.A.~Petrov,
Phys.\ Lett.\ B486, 418 (2000).

\bibitem{e791y}
E.M. Aitala {\it et al.}, E791 Collaboration, Phys. Rev. Lett. 83, 32 
(1999).

\bibitem{focusy}
J.M. Link {\it et al.}, FOCUS Collaboration, Phys. Lett. B485, 62 (2000).

\bibitem{cleoy}
D. Cronin-Hennessy {\it et al.}, CLEO Collaboration, hep-ex/0102006.

\bibitem{belley}
K. Abe {\it et al.}, BELLE Collaboration, BELLE-CONF-0131.

\bibitem{babary}
B.~Aubert {\it et al.}, BaBar Collaboration, hep-ex/0109008.

\bibitem{Godang:2000yd}
R.~Godang {\it et al.},  CLEO Collaboration,
Phys.\ Rev.\ Lett.\  84, 5038 (2000).

\bibitem{brand}
G.~Brandenburg {\it et al.}, CLEO Collaboration,
Phys.\ Rev.\ Lett.\ 87, 071802 (2001).

\bibitem{DmixSL}
E.M.~Aitala {\it et al.}, E791 Collaboration, Phys.\ Rev.\ D57, 13 
(1998).

\bibitem{Silva:2000bd}
J.~P.~Silva and A.~Soffer,
Phys.\ Rev.\ D61, 112001 (2000).

\bibitem{Gronau:2001nr}
M.~Gronau, Y.~Grossman and J.~L.~Rosner,
Phys.\ Lett.\ B508, 37 (2001).

\bibitem{nelson}
H.N. Nelson, hep-ex/9908021.

\bibitem{Bigi:2000wn}
I. Bigi and N. Uraltsev, Nucl. Phys. B592, 92 (2000).

\bibitem{Buccella:1996uy} F.~Buccella, M.~Lusignoli and A.~Pugliese,
Phys.\ Lett.\ B379, 249 (1996).

\bibitem{Golowich:1998pz}
E.~Golowich and A.A.~Petrov, Phys.\ Lett.\ B427, 172 (1998).

\bibitem{wolf}
L. Wolfenstein, Phys.\ Lett.\ B164, 170 (1985).

\bibitem{cnp}
P. Colangelo, G. Nardulli and N. Paver,  Phys.\ Lett.\ B242, 71 (1990).

\bibitem{kaeding}
T.A. Kaeding,  Phys. Lett. B357, 151 (1995).

\bibitem{apreview}
A.A.~Petrov, hep-ph/0009160:
E.~Golowich, ``$CP$ violation in the
charm sector,'' {\it Workshop on CP Violation,
Adelaide, Australia, July 3-8, 1998}.

\bibitem{Manohar:1984md}
A.~Manohar and H.~Georgi,
Nucl.\ Phys.\ B 234, 189 (1984).

\bibitem{PDG}
D.E.~Groom {\it et al.}, Particle Data Group, Eur.\ Phys.\ J.\ C15, 1
(2000).

\bibitem{Falk:1999ts}
A.F.~Falk, Y.~Nir and A.A.~Petrov, JHEP 12, 019 (1999).

\bibitem{MJS}
M.J. Savage, Phys.\ Lett.\ B257, 414 (1991).

\bibitem{LSW}
Z.~Ligeti, I.W.~Stewart and M.B.~Wise, Phys.\ Lett.\ B420, 359 (1998).

\bibitem{NP}
S. Nussinov and M.V. Purohit, hep-ph/0108272.

\bibitem{Golowich:1981yg} E.~Golowich,
Phys.\ Rev.\ D24, 676 (1981).

\bibitem{Gronau:1999zt}
M.~Gronau, Phys.\ Rev.\ Lett. 83, 4005 (1999).


\end{thebibliography}
\end{document}